\documentclass[onecolumn]{article}
\usepackage[english]{babel}

\usepackage[utf8]{inputenc}
\usepackage[pdftex]{graphicx}
\usepackage{tabularx}
\usepackage{authblk}
\usepackage{dcolumn}
\usepackage{amsmath}
\usepackage{longtable}
\usepackage{tabularx}
\usepackage{setspace}
\usepackage{multirow}
\usepackage{mathtools}
\usepackage{tikz}
\usepackage{lscape}
\usepackage{amssymb}
\usepackage{url}
\usepackage{arydshln}
\usepackage{lscape}
\usepackage[misc]{ifsym}
\usepackage{apalike}

\newcommand{\beginsupplement}{%
        \setcounter{table}{0}
        \renewcommand{\thetable}{S\arabic{table}}%
        \setcounter{figure}{0}
        \renewcommand{\thefigure}{S\arabic{figure}}%
        \setcounter{equation}{0}
        \renewcommand{\theequation}{S\arabic{equation}}%
     }

\usepackage{csquotes}
\MakeOuterQuote{"}
\usepackage{lscape}

\usepackage[resetlabels]{multibib}
\newcites{latex}{Supplementary references}%  \citelatex, \nocitelatex, ...

\usepackage[font=small]{caption}

\RequirePackage[twoside,letterpaper,includeheadfoot,layoutsize={8.125in,10.875in},layouthoffset=0.1875in,layoutvoffset=0.0625in,left=38.5pt,right=43pt,top=43pt,bottom=32pt,headheight=0pt,headsep=10pt,footskip=25pt]{geometry}
\title{The rising of collective forgetting and cultural selectivity in inventors and physicists communities}

\author[1,2,3, \Letter]{\small Cristian Candia}

\author[1,2]{Brian Uzzi}

\affil[1]{\small Kellogg School of Management, Northwestern University, Evanston, IL 60208}
\affil[2]{Northwestern Institute on Complex Systems (NICO), Northwestern University, Evanston, IL 60208}

\affil[3]{Centro de Investigación en Complejidad Social (CICS), Facultad de Gobierno, Universidad del Desarrollo, Las Condes 7550000, Chile}

\affil[ \Letter ]{cristian.candia@kellogg.northwestern.edu}

\date{\today}    
\begin{document}

%\twocolumn[\begin{@twocolumnfalse}
\maketitle

\begin{abstract}

How long until this paper is forgotten? Collective forgetting is the process by which the attention received by cultural pieces decays as time passes. Recent work modeled this decay as the result of two different processes, one linked to communicative memory --memories sustained by human communication-- and cultural memory --memories sustained by the physical recording of content. Yet, little is known on how the collective forgetting dynamic changes over time. Are older cultural pieces forgotten at a lower rate than newer ones? Here, we study the temporal changes of collective memory and attention by focusing on two knowledge communities: inventors and physicists. We use data on patents from the United States Patent and Trademark Office (USPTO) and physics papers published in the American Physical Society (APS) to quantify how collective forgetting has changed over time. The model enables us to distinguish between two branches of forgetting. One branch is short-lived, going directly from communicative memory to oblivion. The other one is long-lived going from communicative to cultural memory and then to oblivion. The data analysis shows an increasing forgetting rate for both communities as the information grows. Furthermore, these knowledge communities seem to be increasing their selectivity at storing valuable cultural pieces in their cultural memory. These findings provide empirical confirmation on the forgetting as an annulment hypothesis and show that knowledge communities can effectively slow down the rising of collective forgetting at improving their cultural selectivity.
\end{abstract}

% \keywords{Collective Attention, Collective Memory, Collective Forgetting, Mathematical Modeling, Computational Social Science}

\section*{Introduction}

Collective attention is central for decision making and the spread of ideas and information in social networks. Most of the progress in this field has been made in small laboratories studies and the theoretical literature of attention economics \cite{Falkinger2007,Simon1971}. Psychologist \cite{Kahneman1973,Pashler1998}, economists \cite{Camerer2003}, and marketing researchers \cite{Pieters1999,Dukas2004,Reis2006} have studied it at the individual and small group level \cite{Wu2007}. Yet, research on collective attention is short on quantitative models that would enable us to connect it empirically to large-scale data, and answer questions such as what are the mechanisms that drive the temporal changes of collective attention and how different communities react to those changes?

Recently, the Computational Social Science field research have relied on a data driven approach to understand and model collective attention. Empirical literature on collective attention models the adoption and diffusion of cultural content as a combination of two processes \cite{Candia2018,Higham2017,Higham20172,Valverde2007,Csardi2007,Golosovsky2012,DorogovtsevMendes2000,Wang2013} preferential attachment and temporal decay or obsolescence. Preferential attachment, \cite{Barabasi1999,Albert2002} or cumulative advantage \cite{Price1976,Yule1925,Merton1988}, refers to a process in which attention begets attention. Think of two inventors’ patents, one with 1,000 citations and another one with ten at a given time $t$. The probability that the first paper receives a new citation at the time $t+1$ is larger than that the second one does, simply because more people already know about it, which translates into a growth of current attention. However, this growth competes with the temporal decay. Think about the current collective attention of quantum gravity articles, the top-10 most listened papers published in 2010 cannot compete with the top-10 most cited papers published in 2018. Even though they achieved the same accomplishments (the top-10 most cited papers in their cohort), the level of current collective attention is higher for 2018’s papers than for past papers. This fact is due to the novelty of scientific articles (and almost every type of cultural productions) decreases with time, because of habituation or competition from other new articles given the limited collective attention capacity, thus, the collective attention that they receive also decreases \cite{Falkinger2007,Wu2007}. Therefore, the dynamics of collective attention is driven by two mechanisms: the growth in the number of people that attend to a given cultural piece (preferential attachment or cumulative advantage mechanisms) and the habituation or competition from other cultural pieces that makes the same cultural piece less likely to be attractive as time goes on (temporal decay or obsolescence). 

Formally, it has been shown that collective attention, A(t), can be modeled as a separable function of two components \cite{Candia2018,Higham2017,Higham20172,Valverde2007,Csardi2007}, and it takes the form: $A(t)=c(k) \times S(t)$, where $c(k)$ captures the effects of preferential attachment and $S(t)$ captures the temporal decay. While the preferential attachment mechanisms is well understood, these models have described the temporal decay mechanism (mostly on paper and patent citations datasets) using exponential and log-normal functions \cite{Higham2017,Higham20172,Wang2013}. Yet, recent research \cite{Candia2018} proposes a universal function that describes the temporal dimension as a bi-exponential function, which uncovers the communicative and cultural memory nature of the temporal decay of collective attention. 

To understand the communicative and cultural memory mechanisms of collective attention decay, \cite{Coman2019} provided a brilliant example. Days after the  September 11th attacks, New Yorkers gathered together in small groups of close friends to debate about the events and their experiences \cite{Mehl2003}. They communicated with each other as the easiest way to share information and to deal with the trauma. Following the immediate consequences of the 9-11 events, newspaper articles, makeshift memorials, books on the subject, and official commemorations served as recurrent reminders of the tragedy. Both the communicative acts occurred right after the attacks (i.e., communicative memories) and the physical artifacts that served as reminders of the attacks (i.e., cultural memories) impacted the collective memories of New Yorkers. In other words, the attack catches the attention of small groups of people, who may further pass it on to others, therefore a positive-reinforcement effect sets in such that the more popular (traumatic in this case) the story becomes, the faster it spreads.

After accounting by preferential attachment effects, it has been shown that there might be a regular pattern of decay of collective attention that characterizes a wide range of cultural products, from academic papers describing the quantum entanglement to Luis Fonsi’s “Despasito” song. This decay pattern is circumscribed by the two factors illustrated in the September 11 example: the communicative interactions occurring right after the release of a cultural product and the physical records that are subsequently established. Thus, two fundamental mechanisms in the temporal dimension of collective attention are unveiled by using a computational social science approach and Big Data Methods.

This literature corpus suggests that communities experience two temporal phases of memory, characterized by different levels of attention: an initial phase of high attention, followed by a longer and slower phase of forgetting. Therefore, it is possible to state that the existence of three fundamental mechanisms behind the collective attention dynamics: Preferential Attachment, Communicative Interactions (associated to communicative memory, i.e., oral transmission of information), and Accessing Records (associated to cultural memory, i.e., physical recording of information). Thus, collective memory \cite{Halbwachs1997,Assmann2008,Assmann2011,Wertsch2002,Goldhammer1998,Zaromb201,Roediger2014,Roediger2009,Rubin2014,Garcia-Gavilanes2017,Candia2018,erll2011memory} can provide insights for building mechanistic models of the decay of collective attention.

\subsection*{Collective Memory}

Collective memories are all the memories, knowledge, and information sustained by communities (as large as all the speakers of a language, or as small as a family) that at the same time shape communities’ identities \cite{Hirst2008}. Communities sustain collective memories by two mechanisms communicative memory and cultural memory. For decades, scholars studying collective memory have advanced a large number of definitions, mechanisms, and processes, helping characterize different forms of collective memory and the key features that contribute to their preservation \cite{Hirst2018}.

To illustrate how collective attention and collective memory are associated, let’s consider the following example. In a college library, we can heuristically identify three different sections by the number of people consulting books (i.e., levels of collective attention): i) The front part houses the most recent textbooks and other material discussed in classes. On demand, new versions of textbooks come out periodically, that usually include reviews and new pieces of related knowledge. ii) The back part contains older books, consulted by a specific audience, or probably a recent new version of them came out with revisions and new related knowledge, thus replacing them in the front part. iii) The sub-basement: The books that live here are in the catalogue, but they are barely being consulted because a new paradigm came out or just because people do not have interest in them, although they have the potential to become relevant if something changes in the system. 

The metaphor of the college library shed lights of the difference and relationship between actively remembered knowledge, or in Assmann’s terms \cite{Assmann2008} the canon - "actively circulated memory that keeps the past present" (p.98), and the archive - "passively stored memory that preserves the past" (p.98). Memory studies’ scholars have extensively argued that culture is inextricably linked to memory \cite{Halbwachs1992,Halbwachs1997,Assmann1995} and its counterpart, forgetting \cite{Assmann2008}. Furthermore, \cite{Assmann1995} defines the existence of two modes of cultural memory "... the mode of potentiality of the archive whose accumulated texts, images, and rules of conduct act as a total horizon, and ... the mode of actuality, whereby each contemporary context puts the objectivized meaning into its own perspective, giving it its own relevance." (p.130). Potentiality is the existence of a record (e.g., old book in a library basement), and actuality being the attention received by that record when it becomes relevant to the community.

We can reasonably assume that the first level of attention, described in the library example, is mainly shaped by communicative acts -- e.g., classes -- that focus the attention of specific communities in specific pieces of knowledge \cite{Candia2018}. Cultural pieces that receive most of the collective attention live in a mode of actuality \cite{Assmann1995}, form part of the canon \cite{Assmann2008}, and their current interpretation is part of the community's identity. 

The second level of attention is mainly shaped by accessing records -- e.g., literature searching -- and the access can be related to both top-down and bottom-up mechanisms. Top-down mechanisms depend on the existing context of the community \cite{Hirst2018} and how it contributes to the retention and formation of collective memories such as, familiarity \cite{Roediger2016,Rubin1995}, narrative templates \cite{Wertsch2002,Hammack2011}, and cultural attractors \cite{Sperber2004,Buskell2017,Richerson2005}. For instance, familiarity increases the memorability of events, even causing false memories, like that of people identifying Alexander Hamilton as a U.S. president \cite{Roediger2016}. Narrative templates, which are schemata that people use to describe multiple historical events, can also shape memories, such as the memory of Russian exceptionalism that emerges from the narrative template of invasion, near defeat, and heroic triumph \cite{Wertsch2002}. Cultural attractors, such as repetitive children’s songs or count-out rhymes, can increase the preservation of memories across generations \cite{Rubin1995}. Bottom-up mechanisms depend on micro-level psychological processes that drive social outcomes \cite{Hirst2018} such as, retrieval-induced forgetting \cite{Storm2012,Garcia-Bajos2009,Cuc2007}, and social affinities, which increase the mnemonic power of conversations \cite{Echterhoff2009,Coman2015,Coman2014,Stone2010}. For instance, people share realities with those whom they perceive as belonging to their own group \cite{Coman2015}.  Therefore, we can say those cultural products live in the transition between the mode of actuality and the mode of potentiality.

The third level of attention is very close to zero, it is reasonable to argue that those books, in Assmann's words, live in a mode of potentiality \cite{Assmann1995}. They are stored in an archive, are not part of the current conversation of the community, but they can be retrieved if the community decide to remember those pieces. 

Thus, remembering and forgetting are dynamic. As \cite{Assmann2008} argues, "[e]lements of the canon can ... recede into the archive, while elements of the archive may be recovered and reclaimed for the canon." (p. 104). Therefore, temporal changes and dynamics are the main focus of this project.

\subsection*{A Computational Social Science approach}

Halbwachs in 1925 stated, “the idea of an individual memory, absolutely separate from social memory, is an abstraction almost devoid of meaning” \cite{Halbwachs1992,Connerton1989}. But the fact that the individual memory cannot be conceptualized and studied without using its social context does not necessarily imply the reverse \cite{Kansteiner2002}. It means that collective memory, although social, can only be imagined and accessed through its manifestation in individuals. Therefore, it is necessary to differentiate between autobiographical memory and collective memory. For lack of such differentiation, many inquiries into collective memories have perceived and conceptualized collective memory exclusively in terms of the psychological and emotional dynamics of individual remembering \cite{Kansteiner2002}. The classic literature about collective memory has done enormous progress focused on the individuals' memories \cite{Hill2013,Roediger2014,Roediger2009,Rubin1996,Rubin1998,Rubin2014,Wixted1997,Zaromb2014}. Yet, Jeffrey Olick, cited by \cite{Kansteiner2002}, proposed a distinction between "collected" and "collective" memory. A collected memory is the classic aggregation of individual data. Still, the collective memory does not have the same dynamics \cite{erll2011memory,Kansteiner2002}; therefore, it is necessary to find another suitable empirical method to study the collective memory. 

This work uses a Computational Social Science approach that relies on data-sets that have the advantage of proxying collective memory through individuals' manifestations and capturing at the same time, the underlying context of the studied communities. Particularly, citation data carries the collective behavior of people in their respective social contexts at being an expression of their individual decision-making process, which considers each subtlety in each context, on citing a paper or a patent. We can think of them as cultural revealed preferences.

Literature on Computational Social Science has used Big Data methods to study the actuality mode \cite{Assmann1995} of memories. Concretely, the area has focused its effort on the consumption of cultural content from Wikipedia page views \cite{Garcia-Gavilanes2017,Yu2016,jara2019medium,Skiena2014,Kanhabua2014,Candia2018} to paper and patent citations \cite{Wang2013,Uzzi2013,Mukherjee2017,Higham2017,Higham20172,Candia2018}, and on the effects of accomplishments, technology, language, and triggers in the dynamics of collective memory and attention. For instance, changes in communication technologies, such as the rise of the printing press, radio, and television have also been shown to affect attention since they correlate with changes in the occupations of the people entering biographical records \cite{jara2019medium}. Historical figures born in countries with languages that are often translated to other languages receive more online attention than comparable historical figures born in areas with less frequently translated languages \cite{Ronen2014}. The edits and attention received by events in Wikipedia have also been seen to increase with related exogenous events \cite{Kanhabua2014,Garcia-Gavilanes2017}, such as human-made and natural disasters, terrorism, accidents, and during anniversaries or commemorative events \cite{Ferron2014}. Moreover, the online attention received by past sports figures --a measure of their prevalence in current memory-- correlates with performance, after discounted for age, \cite{Yu2016,Yucesoy2016}, meaning that attention and memorability correlate with merit in athletic activities. The online and offline metrics aforementioned are indirect measures of collective memory and attention. They measure attention spillovers that result in online searches or references. The idea is that cultural pieces that are being talked about and are of heightened interest, and hence, lead people to consult various data sources. When these cultural products move away from communicative memory, they lose the intense attention they initially had.

Thus, the Computational Social Science approach works close to Assmann’s definition of collective memory \cite{Assmann1995,Assmann2011}, which focuses on the cultural products that communities or groups of people remember. Assmann focused on long-lived inter-generational memories, yet, he distinguishes between modes of potentiality and actuality. 

\section*{Method}

\subsection*{Model}

We model the collective attention received by cultural pieces by assuming that the collective attention size occupied by comparable cultural pieces for all time $t$ is $S(t)=u(t)+v(t)$. $S$ is the sum of the collective attention size on communicative memory ($u$) and cultural memory ($v$). We note that the decay of the collective attention size of a particular cultural piece is entirely random in time so it is impossible to predict when a particular cultural piece will be remembered. However, the collective attention size ($S(t)$) for a particular cultural piece is equally likely to decay at any instant in time. Therefore, given a set of comparable cultural pieces (discounted by preferential attachment and having the same age), the number of remembering acts $-dS$ expected to occur in a small interval of time $dt$ is proportional to the size of collective attention $S$, that is $\frac{-dS}{dt}=\propto S \Longrightarrow \frac{-dS}{S}=\propto dt\label{eq_gen}$.

\begin{figure}[!h]
 \centering 
  \includegraphics[width=0.9\textwidth]{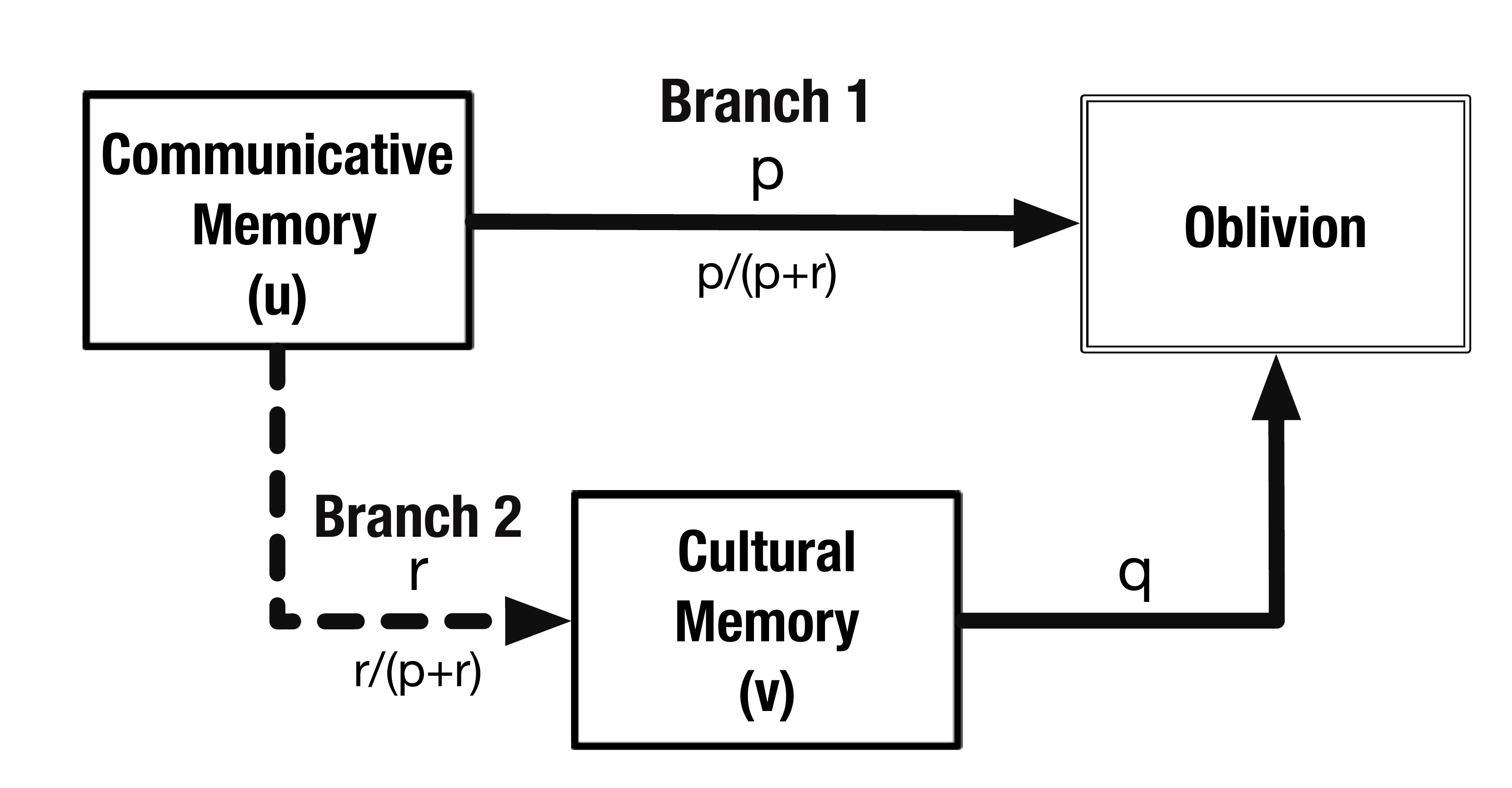}
\caption{Collective forgetting model based on collective memory. The single stroke boxes indicate that both communicative and cultural memory are transient states of the model. The double stroke box indicates that oblivion is an absorbing state of the model. Branch 1 is the path from communicative memory to oblivion. Branch 2 is the path from communicative memory to cultural memory to oblivion. The parameter $r$ quantifies the rate at which communities include cultural pieces from communicative memory to cultural memory. Parameters $p$ and $q$ represent the forgetting rates of communicative and cultural memory respectively. An $r/(p+r)$ share of cultural pieces are preserved longer in the cultural memory because they receive more attention, while a $p/(p+r)$ share of cultural pieces are forgotten. Communities can effectively change the rate ($r$) at which they store cultural pieces in cultural memory (dashed arrow).}
 \label{fig_model}
\end{figure}

Thus, the expected decay of collective attention size, $-dS/S$, is proportional to an increment of time, $dt$, where the negative sign indicates that the collective attention size decreases as time increases. Considering that $S=u+v$, with initial conditions $u(0)=N$ and $v(0)=0$, the solution of this coupled differential equation system \cite{Candia2018} depicted in Figure \ref{fig_model} is:
\begin{equation}
u(t) = Ne^{-(p+r)t}, \qquad v(t) = N\frac{r}{p+r-q} \left[ e^{-qt}-e^{-(p+r)t} \right], \label{cultural_model}%\\
\end{equation}
where $N$ is the total size of collective attention at time $t=0$, $p$ is the decay rate from communicative memory to oblivion, $q$ is the decay rate from cultural memory to oblivion, and $r$ is the coupling rate between communicative and cultural memory, it means, the rate at which communities include cultural pieces in their cultural memory.% (Fig. \ref{fig_main} E, F, and I for APS papers and G, H, and J for USPTO patents). 

Finally, the forgetting rate for the system is defined as $\lambda= q(r+p)/(r+q)$ (See supplementary material \ref{SM_lambda} for the derivation). The branching ratios are straightforward defined for the attention that goes from communicative acts to oblivion as $B_1=p/(p+r)$, and for the attention that goes from communicative acts to records consultation as $B_2=r/(p+r)$. They represent the relative probability for each branch decay, in other words, the probability that cultural content goes directly to oblivion, $B_1$, or to oblivion through cultural memory $B_2$.

\subsection*{Data description}

We use time series data on scientific articles and patents to investigate the forgetting dynamics of inventors and physicists communities. For patents, the  United States Patent and Trademark Office (USPTO) corpus is used. It contains data on $n=1,681,690$ patents granted in United States from 1973 to 2005 classified in six different categories Chemical (CAT 1), Computers $\&$ Computation (CAT 2), Drugs $\&$ Medical (CAT 3), Electrical $\&$ Electronic (CAT 4), Mechanical (CAT 5) and Others (CAT 6). For articles, we use the American Physical Society (APS) corpus, which contains data about $n=485,105$ physics articles from twelve different journals\footnote{Some of these journals do not exist anymore, some others are too small in terms of the number of publications, and one of them only publishes reviews, which follow a completely different citation dynamics.}, between 1896 and 2016. Both data-sets have been validated and studied before in the literature \cite{Candia2018,Wang2013,Higham20172,Jaffe1993,Sinatra2016,Shen2014,Higham20172}.

The analysis is restricted to USPTO patents granted from 1976 to 1995 in all categories except Drugs and Medical, and to scientific articles published in PRA, PRB, PRC, PRD, PRE, PRL, and PR from 1950 to 1999.  The data restrictions are based on two considerations: i) We discarded the Drugs $\&$ Medical category from USPTO patents because the data has missing entries for this category between 1990 and 1995. ii) A prospective approach is used to analyze citation data (see supplementary figure \ref{approaches}), which means that the received citations are tracked to the future for each cohort of published or granted documents. Therefore, this approach demands a time window of at least five years after the transition time\footnote{The transition time is defined as the time at which cultural memory overcomes communicative memory \cite{Candia2018}, that is for papers $t_c \approx 9$ years and for patents and $t_c \approx 5$ years.} ($t_c$) to have reliable parameters from the forgetting model.

\subsection*{Panel Data Analysis}

Panel data analysis (PDA) is broadly used in social science to analyze multi-dimensional data involving observations over multiple periods (cohorts) over the same entities (USPTO categories and APS journals). A pooled model, which assumes that all observations are independent, is first estimated as $y_{it}=\alpha+\beta x_{it}+u_{it}$, where $x_{it}$ is a vector of independent variables, and $u_{it}$ is the error terms. However, the effects of independent variable could be driven by omitted variables, i.e., $cov(x_{it}; u_{it}) \neq 0$. Thus, the panel data structure of the data is exploited to provide evidence on the effects of independent variables beyond and above the entities' invariant characteristics and temporal changes. The within estimation follows the same general structure as the pooled model, but the error term is modeled as $u_{it}=c_i+\lambda_t+v_{it}$, where $c_i$ absorbs all the invariant omitted variables across the grouping index $i$. $\lambda_t$ absorbs all the temporal changes, and, $v_it$ is the unobserved error term. The within estimator that considers just the entities' invariant effects ($c_i$) is called the one-way estimator, whereas the estimation that considers both entities ($c_i$) and temporal ($\lambda_t$) effects is called the two-ways estimator.

\subsection*{Procedure}

For both patents and papers, two time-series were built. One for the new number of citations, which is built as the number of citations obtained in a one year time window \cite{Higham2017,Higham20172,Candia2018}. The second time series is for the accumulated citations obtained up to a given time. 

Next, a 'prospective approach' (see supplementary figure \ref{approaches}) is implemented \cite{Glanzel2004,Bouabid2011,YIN2017608}, which focuses on how specific cohorts of cultural production (e.g., all paper published in PRA in 1990) acquire new citations over time. The prospective approach is used because there is no mathematical difference with the retrospective approach \cite{YIN2017608}. Moreover, empirically speaking, is more simple to work with the prospective approach than the retrospective approach \cite{Glanzel2004,Bouabid2011}.

To isolate the temporal dimension of collective forgetting, all cultural pieces (papers or patents) are grouped according to their logarithmic number of accumulated citations. Thus, we have a time series for each cohort of cultural pieces in each group of accumulated citations.

Then, a two panels are built using journals (APS) and categories (USPTO) as grouping index and the year of the cohort as the temporal index. Thus, pooled, within one-way, within two-ways, and first differences models are estimated.

Finally, two groups of documents for each community are defined: i) All documents. This group includes all documents published and granted of each cohort. ii) The top-cited documents. This group includes the top $15\%$ of the most cited documents between $t_c\pm 1$ years after release. In other words, the top-cited groups are the most cited patents ($t_c=5$) between the fourth and the sixth year after granted; and the most cited papers ($t_c=9$) between the eighth and tenth year after publication. 

\subsection*{Limitations}

Unfortunately, the aggregate approaches cannot distinguish between different forms of memory or attention loss, such as interference, suppression, or inhibition. They only provide an aggregate picture of the attention lost through all of these channels.

\section*{Results}
This paper focuses on the temporal dimension of collective memory and attention, $S(t)$, using a model grounded on the collective memory literature, where cultural memory and communicative memory co-exist, but decay at different rates \cite{Candia2018}. Figure \ref{fig_model} shows a diagram of the model, where all the collective attention of cultural pieces created in a certain year, is due the communicative memory. Then, it decays following two pathways towards forgetting (oblivion). Branch 1 of forgetting goes from communicative memory to oblivion, and branch 2 goes from communicative to cultural memory and then to oblivion. Thus, this two-step forgetting model enables us to further explore the dynamics of forgetting.

\begin{figure}[!h]
 \centering 
  \includegraphics[width=0.95\textwidth]{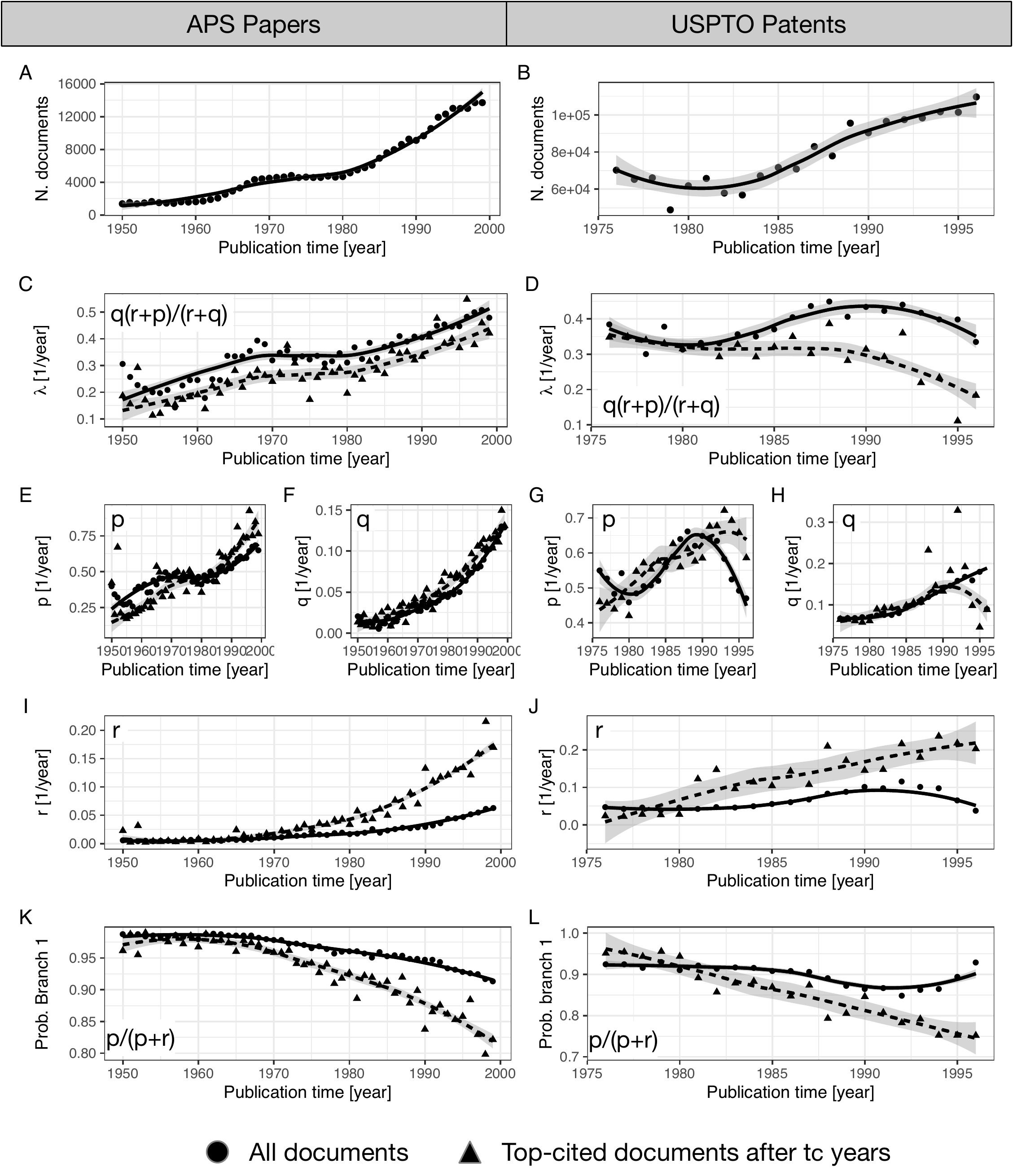}
\caption{Attention decay for APS (left column) and USPTO communities (right column). Solid lines represent all aggregated documents, and dashed lines represent the top-cited 15 $\%$ of documents after $t_c$ years. \textbf{A} Total number of published papers. \textbf{B} Total number of granted patents. \textbf{C and D} Forgetting rate ($\lambda$). \textbf{E and G} Decay ratio from communicative memory to oblivion ($p$).  \textbf{F and H} Decay ratio from cultural memory to oblivion ($q$). \textbf{I and J}  Coupling rate from communicative memory to cultural memory ($r$). \textbf{K and L} Probability of branch 1. The share of documents that are directly forgotten, i.e, that go from communicative memory to oblivion.}
 \label{fig_main}
\end{figure}

How do physicists and innovators communities forget over time? Model parameters ($p$, $q$, and $r$), and the forgetting rate of the American Physical Society (APS) and the United States Patent and Trademark Office (USPTO) communities ($\lambda= q(r+p)/(r+q)$) follow an increasing trend over the years (Figs. \ref{fig_main}C-J). These trends suggest, on average, an accelerating forgetting. But, why these communities forget quicker over time? We observe that the number of scientific articles published by the APS and the number of granted patents by the USPTO also increase over time (Fig. \ref{fig_main}A and B). The similarity of these trends is in line with theoretical literature that suggests a relationship between the increasing amount of information and forgetting rates, a phenomenon called forgetting as an annulment \cite{Connerton2008}. 

The forgetting as annulment hypothesis is tested for both APS (Table \ref{table_papers}) and USPTO (Table \ref{table_patents}) communities and their sub-communities. The sub-communities are defined as the different knowledge fields that different APS journals and USPTO categories represent. A significant and positive effect is observed for each specification in both communities, providing evidence on the excess of information induce forgetting. The effect survives above and beyond unobserved variables captured by the entities' fixed effects (Models 3 and 4) and both entities fixed effects and temporal fixed-effects (Models 5 and 6). Moreover, given that some of the specification exhibit serial correlation in the errors ($u_{it}$), the coefficients using robust standard errors are reported for each specification. The \cite{driscoll1998consistent} method (Models 2, 4, and 6) was used, which accounts for heteroskedasticity, auto-correlation, and cross-sectional dependence of the errors. Besides, we observe no serial correlation in the difference of the errors ($\Delta u_{it}$); hence, a first difference estimator is computed (Models 6). This model also includes a lag for the number of documents and a lag for the forgetting rate. The first difference estimator also supports forgetting as an annulment hypothesis in both communities. We conducted the Durbin–Wu–Hausman test \cite{Greene2003} and the null hypothesis was rejected in all specifications (p-value$< 0.01$), leading to select the within estimator instead of the random estimator for the analysis.

But, do the communities respond to the forgetting as an annulment? Evidence suggest that the communities fight back the forgetting as an annulment. Indeed, the forgetting rate, $\lambda$, is always lower for top-cited cultural pieces\footnote{Defined as the top $15\%$ of the most cited documents between $t_c\pm 1$ years after release} (dashed lines in Figs. \ref{fig_main}C and D) compared to all documents (solid lines in Figs. \ref{fig_main}C and D). Then, which mechanism does drive the observed smaller forgetting rate? The decay rates of communicative memory ($p$) and cultural memory ($q$) are statistically equivalent for both groups of documents (Figs. \ref{fig_main}E, F, G, and Q). This equivalence suggests that the communities are not able (or at least it is complicated) to manipulate or intervene on these rates; hence, the differences in the forgetting rate ($\lambda$) between the two groups are not due to the individual forgetting rates ($p$ and $q$). Nevertheless, the APS and USPTO communities increase faster its coupling rate ($r$) for top-cited cultural pieces (dashed line Figs. \ref{fig_main}I and J) compared to all produced cultural pieces (solid line Figs. \ref{fig_main}I and J). Finally, as a consequence of the differences in the coupling parameter, the probability that scientific articles and patents decay directly to oblivion (branch 1, $B_1= p/(p+r)$) is smaller for top-cited cultural pieces (dashed lines in Figs. \ref{fig_main}K and L) than all produced cultural pieces (solid lines in Figs. \ref{fig_main}K and L).

% Table created by stargazer v.5.2.2 by Marek Hlavac, Harvard University. E-mail: hlavac at fas.harvard.edu
% Date and time: Thu, Aug 13, 2020 - 22:44:48
\begin{table}[!htbp] \centering 
  \caption{Regression models for APS papers.} 
  \label{table_papers} 
\tiny 
\begin{tabular}{@{\extracolsep{-10pt}}lccccccc} 
\\[-1.8ex]\hline 
\hline \\[-1.8ex] 
 & \multicolumn{7}{c}{\textit{Dependent variable:}} \\ 
\cline{2-8} 
\\[-1.8ex] & \multicolumn{6}{c}{Forgetting Rate}  & Forgetting Rate  (log) \\ 
\\[-1.8ex] & \textit{panel} & \textit{coefficient} & \textit{panel} & \textit{coefficient} & \textit{panel} & \textit{coefficient} & \textit{panel} \\ 
 & \textit{linear} & \textit{test} & \textit{linear} & \textit{test} & \textit{linear} & \textit{test} & \textit{linear} \\ 
\\[-1.8ex] & (1) & (2) & (3) & (4) & (5) & (6) & (7)\\ 
\hline \\[-1.8ex] 
 N. papers & 0.52$^{***}$ & 0.52$^{***}$ & 0.75$^{***}$ & 0.75$^{***}$ & 0.17$^{**}$ & 0.17$^{***}$ &  \\ 
  & (0.06) & (0.07) & (0.06) & (0.09) & (0.08) & (0.06) &  \\ 
  & & & & & & & \\ 
 N. papers (log) &  &  &  &  &  &  & 0.30$^{**}$ \\ 
  &  &  &  &  &  &  & (0.12) \\ 
  & & & & & & & \\ 
 Lag N. papers (log) &  &  &  &  &  &  & 0.12 \\ 
  &  &  &  &  &  &  & (0.11) \\ 
  & & & & & & & \\ 
 Lag Forgetting Rate ($\lambda$) &  &  &  &  &  &  & $-$0.48$^{***}$ \\ 
  &  &  &  &  &  &  & (0.07) \\ 
  & & & & & & & \\ 
 Constant & 0.29$^{***}$ & 0.29$^{***}$ &  &  &  &  &  \\ 
  & (0.01) & (0.01) &  &  &  &  &  \\ 
  & & & & & & & \\ 
\hline \\[-1.8ex] 
Effects: & Pooling & SCC & Within & SCC & Within & SCC & First differences \\ 
 &  &  & One-way &  & Two-ways &  &  \\ 
Observations & 189 &  & 189 &  & 189 &  & 175 \\ 
R$^{2}$ & 0.30 &  & 0.49 &  & 0.04 &  & 0.24 \\ 
Adjusted R$^{2}$ & 0.30 &  & 0.47 &  & $-$0.37 &  & 0.23 \\ 
\hline 
\hline \\[-1.8ex] 
\textit{Note:}  & \multicolumn{7}{l}{$^{*}$p$<$0.1; $^{**}$p$<$0.05; $^{***}$p$<$0.01} \\ 
 & \multicolumn{7}{l}{Note that the number of papers is scaled by 100,000 and for first differences} \\ 
 & \multicolumn{7}{l}{model the data was logarithmized.} \\ 
 & \multicolumn{7}{l}{SCC: \cite{driscoll1998consistent} method. Robust errors for heteroskedasticity,} \\ 
 & \multicolumn{7}{l}{autocorrelation, and cross-sectional dependence.} \\ 
\end{tabular} 
\end{table} 

% Table created by stargazer v.5.2.2 by Marek Hlavac, Harvard University. E-mail: hlavac at fas.harvard.edu
% Date and time: Thu, Aug 13, 2020 - 21:54:26
\begin{table}[!htbp] \centering 
  \caption{Regression models for USPTO patents.} 
  \label{table_patents} 
\tiny 
\begin{tabular}{@{\extracolsep{-10pt}}lccccccc} 
\\[-1.8ex]\hline 
\hline \\[-1.8ex] 
 & \multicolumn{7}{c}{\textit{Dependent variable:}} \\ 
\cline{2-8} 
\\[-1.8ex] & \multicolumn{6}{c}{Forgetting Rate}  & Forgetting Rate  (log) \\ 
\\[-1.8ex] & \textit{panel} & \textit{coefficient} & \textit{panel} & \textit{coefficient} & \textit{panel} & \textit{coefficient} & \textit{panel} \\ 
 & \textit{linear} & \textit{test} & \textit{linear} & \textit{test} & \textit{linear} & \textit{test} & \textit{linear} \\ 
\\[-1.8ex] & (1) & (2) & (3) & (4) & (5) & (6) & (7)\\ 
\hline \\[-1.8ex] 
 N. papers & 0.55$^{***}$ & 0.55$^{***}$ & 1.22$^{***}$ & 1.22$^{***}$ & 0.74$^{**}$ & 0.74$^{***}$ &  \\ 
  & (0.12) & (0.15) & (0.16) & (0.20) & (0.37) & (0.13) &  \\ 
  & & & & & & & \\ 
 N. papers (log) &  &  &  &  &  &  & $-$0.04 \\ 
  &  &  &  &  &  &  & (0.10) \\ 
  & & & & & & & \\ 
 Lag N. papers (log) &  &  &  &  &  &  & 0.22$^{**}$ \\ 
  &  &  &  &  &  &  & (0.09) \\ 
  & & & & & & & \\ 
 Lag Forgetting Rate ($\lambda$) &  &  &  &  &  &  & $-$0.05 \\ 
  &  &  &  &  &  &  & (0.10) \\ 
  & & & & & & & \\ 
 Constant & 0.30$^{***}$ & 0.30$^{***}$ &  &  &  &  &  \\ 
  & (0.02) & (0.03) &  &  &  &  &  \\ 
  & & & & & & & \\ 
\hline \\[-1.8ex] 
Effects: & Pooling & SCC & Within & SCC & Within & SCC & First differences \\ 
 &  &  & One-way &  & Two-ways &  &  \\ 
Observations & 98 &  & 98 &  & 98 &  & 87 \\ 
R$^{2}$ & 0.18 &  & 0.39 &  & 0.05 &  & 0.09 \\ 
Adjusted R$^{2}$ & 0.17 &  & 0.36 &  & $-$0.26 &  & 0.06 \\ 
\hline 
\hline \\[-1.8ex] 
\textit{Note:}  & \multicolumn{7}{l}{$^{*}$p$<$0.1; $^{**}$p$<$0.05; $^{***}$p$<$0.01} \\ 
 & \multicolumn{7}{l}{Note that the number of papers is scaled by 100,000 and for first differences} \\ 
  & \multicolumn{7}{l}{model the data was logarithmized.} \\ 
 & \multicolumn{7}{l}{\cite{driscoll1998consistent} method. Robust errors for heteroskedasticity,} \\ 
 & \multicolumn{7}{l}{autocorrelation, and cross-sectional dependence.} \\ 
\end{tabular} 
\end{table}

\section*{Discussion}

"[A] wealth of information creates a poverty of attention and a need to allocate that attention efficiently among the overabundance of information sources that might consume it," said \cite{Simon1971}. Simon's work has inspired scholars for decades at studying attention economics \cite{davenport2001attention,Lanham2006,Weng2012}. Yet, little is known on the mechanisms that drive the temporal changes of collective attention, and how communities react to those changes. 

Understanding the temporal dynamics of collective attention is a challenge because several factors impact on it. Some of these factors go from psychological phenomena (e.g., moral emotions elicited by particular messages \cite{brady2017emotion}) to socio-structural phenomena (e.g., the communities’ underlying network structures in which the cultural products diffuse \cite{Coman2016}). Nevertheless, the practical applications range from the prediction of performance in sports \cite{Yucesoy2016}, arts \cite{fraiberger2018quantifying}, and science \cite{Wang2013,Higham20172,Candia2018} to the resilience of a country in a traumatic experience (e.g., September 11th) \cite{Mehl2003} or for developing efficient communication strategies to improve policy awareness \cite{cunico2020building}.

Concretely, this study tackles one of the factors that impact collective attention, through the lens of collective memory psychology, and it can be summarised as follow: The excess of information promotes forgetting in knowledge communities, namely technological (inventors) and academic (physicist) communities. This effect was theoretically described by \cite{Connerton2008} under the name of forgetting as annulment. The data analysis supported this hypothesis (Tables \ref{table_papers} and \ref{table_patents}), providing evidence on the relationship between the amount of information produced (number of published or granted documents) and the forgetting rates (estimated using a collective attention model fully grounded on collective memory literature \cite{Candia2018}). 

The forgetting as annulment becomes an important issue when the community's capacity to remember what is worth remembering is affected. A high amount of cultural products and their easy archivalisation create the perfect scenario to forget useful cultural pieces. In the information era, characterized by vast amounts of data, the filtering problem is central to preserve and build knowledge.  Indeed, \cite{Kuhn1970} suggests that paradigm shifts happen when a new knowledge overcomes a former one. With an increasing rate of knowledge production, knowledge communities must adjust their filtering capacities to keep track of which information worth to be remembered and what is not. Thus, storing valuable content in the cultural memory is a potential mechanism for slowing down increasing forgetting rates. It is expected that filtering capacities are more efficient in communities where the production of cultural products depends on objective processes. The data analysis shows that communities of inventors and physicists increase their coupling rates ($r$) for top-cited documents (Figs. \ref{fig_main}I and J). The differences in the coupling rate between all documents and the top-cited documents suggest that both knowledge communities respond to the increasing forgetting. They actively select worth to remember pieces of information and storing them in their cultural memory (Figs. \ref{fig_main}K, and L), where they live for longer in a mode of actuality \cite{Assmann1995}. 

Taken together, these results provide insights into the temporality of collective memory and attention, which could have potential real-world consequences. For instance, the current pandemic crisis related to SARS-CoV-2 is probably the most important world crisis in the last five decades. It is a human tragedy with thousands of deaths around the globe. The pandemic also has an increasing impact on different dimensions ranging from the global economy to mental health. Indeed, some of the effects of epidemics in mental health are post-traumatic stress disorder, depression, substance use disorder, domestic violence, child abuse, and a broad range of other mental and behavioral disorders. The communicative processes following past epidemics could only be sustained for a limited time, a period in which individuals in the affected region probably adjusted their beliefs about, for example, the relevance of mental health. Once the communicative processes related to SARS-CoV-2 will decrease their intensity, and no artifacts serve as reminders, the collective memory and attention related to the severity of the epidemic, will also decay, and with it, the belief that mental health is an issue that needs policy compromises. Understanding the temporal dynamic of collective attention could offer policymakers much needed time to lock in commitment devices at a population level. Indeed, the same rationale applies to natural disasters, climate change, or economic crisis.

% \section*{Code availability}

% In this paper we use the Natural Language Toolkit of Python 3 to clean the data. Then, for topic modeling, we use the package "topicmodels" of R version 3.6.2. For Linguistic Inquiry Word Counting, we use the Quanteda package \cite{benoit2018quanteda} in R 3.6.2.

% A sample of the code for analysing the data and the moral dictionaries used can be found in the supplementary software attachment.

% \section*{Data Availability}
% The necessary data to reproduce this work can be delivered under a reasonable request to the authors. 

\section*{Acknowledgments}
The authors acknowledge to Judit Varga and Cristian Jara-Figueroa for the extremely helpful insights, feedback, and discussions. The authors are also thankful to Carlos Rodriguez-Sickert, Javier Pulgar, Flavio Pinheiro, Victor Landaeta, and Miguel Guevara for their comments and suggestions. The authors also acknowledge the financial support of the Northwestern Institute on Complex Systems and the Kellogg School of Management at the Northwestern University.

\section*{Corresponding Author}
Cristian Candia (cristian.candia@kellogg.northwestern.edu | ccandiav@mit.edu | ccandiav@udd.cl).

\bibliographystyle{apalike}
\bibliography{bibliography}

% \printbibliography

\pagebreak

% \appendix

\setcounter{page}{1}
\section*{Supplementary material}
\beginsupplement

\section{Forgetting rate}\label{SM_lambda}
Let's consider a system with $t=2$ transient states (communicative and cultural memory) and $a=1$ absorbing state (oblivion), where the probability of transitioning from one transient state to another is given by Q and the probability of transitioning from a transient state to the absorbing state is given by R. 

\begin{equation*}
Q=
\begin{bmatrix}
1-r-p & r\\
0 & 1-q \\
\end{bmatrix}
\end{equation*}

and

\begin{equation*}
R=
\begin{bmatrix}
 p\\
 q\\
\end{bmatrix}
\end{equation*}

The canonical form of the transition matrix $P$ of an absorbing Markov chain is given by P.
\begin{equation*}
P=
\begin{bmatrix}
Q & R\\
0 & I
\end{bmatrix}
=
\begin{bmatrix}
1-r-p & r & p\\
0 & 1-q & q\\
0 & 0 & 1
\end{bmatrix}
\end{equation*}

Then, the fundamental matrix, which quantifies the probability of visiting a certain transient state is given by N.
\begin{eqnarray*}
N&=&(I-Q)^{-1}
=
\begin{bmatrix}
r+p & -r\\
0 & q
\end{bmatrix}^{-1}\\
&=&
\frac{1}{q(r+p)}
\begin{bmatrix}
q & r\\
0 & r+p
\end{bmatrix}
\end{eqnarray*}

The N matrix enable us to calculate different properties of a Markov chain. The focus here is the number of steps (time) before cultural productions are absorbed by the Oblivion state starting from both transient states (communicative and cultural memory). 
\begin{eqnarray*}
N\begin{bmatrix}
1\\
1
\end{bmatrix}
&=&
\frac{1}{q(r+p)}
\begin{bmatrix}
r + q\\
r + p
\end{bmatrix}
\end{eqnarray*}

Given the assumption that all the collective attention starts at the communicative memory transient state, the number of stpes before cultural productions starting at the communicative memory are absorbed by the Oblivion state if given by the first component of the previous matrix, $\tau$.  The forgetting rate is defined as $\lambda=1/\tau$.

\begin{eqnarray*}
    \tau&=&\frac{1}{\lambda}=\frac{r+q}{q(r+p)}\\
    \lambda&=&\frac{q(r+p)}{r+q}
\end{eqnarray*}

\clearpage

\section{Prospective approach}\label{SM_prospective}

\begin{figure}[!h]
 \centering 
  \includegraphics[width=0.9\textwidth]{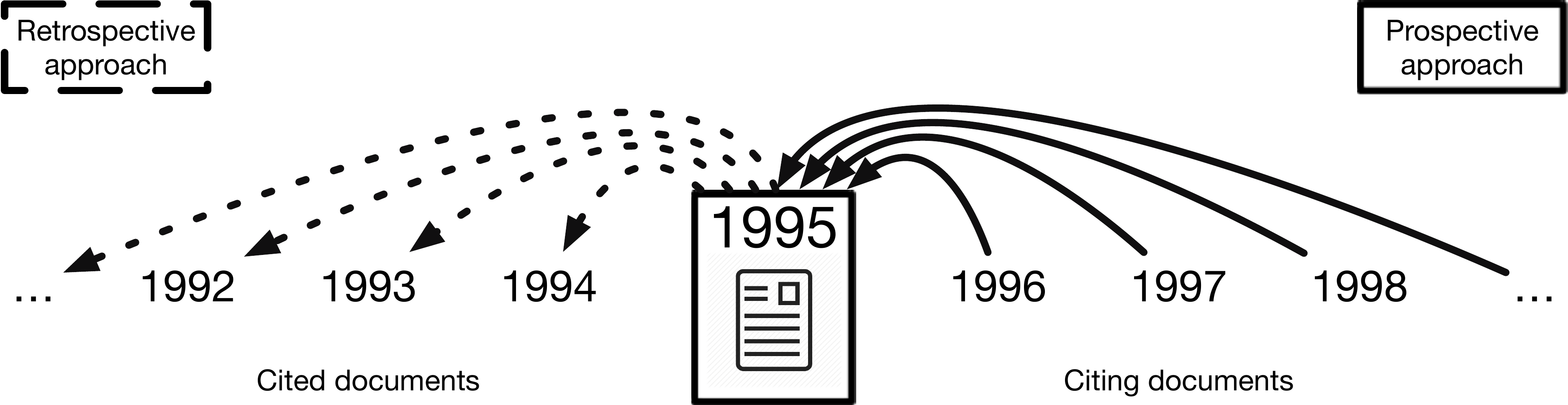}
\caption{Retrospective and prospective approach for tracking citations. In the retrospective approach (dashed arrows) the focus is to track the cited documents of each specific cohort (looking at the past). In the prospective approach (solid arrows) the focus is to track the citing documents of each specific cohort (looking at the future).}
 \label{approaches}
\end{figure}

% \pagebreak
% %\section*{Supplementary References}
% \bibliographystylelatex{apalike}
% \bibliographylatex{references}
% % \printbibliography

\end{document}